\documentclass[a4paper,12pt,twoside]{article}
\usepackage{amsfonts,graphicx} 

\newtheorem{conjecture}{Conjecture}
\begin{document}
\title{\bf A NOTE ON THE  $osp(1|2s)$ 
THERMODYNAMIC BETHE ANSATZ EQUATION 
}
\author{ZENGO TSUBOI\footnote{
E-mail address: ztsuboi@poisson.ms.u-tokyo.ac.jp} \\
{\it Graduate School of Mathematical Sciences, 
University of Tokyo,} \\ 
 {\it Komaba 3-8-1, Meguro-ku, Tokyo 153-8914, Japan}}
\date{}
\maketitle
\begin{abstract}
A Bethe ansatz equation associated with 
the Lie superalgebra $osp(1|2s)$ is studied. 
A thermodynamic Bethe ansatz (TBA) equation is derived by 
the string hypothesis. 
The high temperature limit of the entropy density 
is expressed in terms of the solution of 
the $osp(1|2s)$ version of the $Q$-system. 
In particular for the fundamental representation case, 
we also derive a TBA equation from the $osp(1|2s)$ version of the 
$T$-system and the 
quantum transfer matrix method. This TBA equation is 
identical to the one from the string hypothesis. 
The central charge is expressed by the Rogers dilogarithmic function, 
and identified to s. 
\end{abstract} 
\section{Introduction}
Recently, thermodynamics of quantum integrable spin chains 
 related to superalgebras received attentions. 
In particular, there are several papers \cite{Sch87,Sch92,EK94,BF95,
JKS98,
Fr99,Sa99} on thermodynamic Bethe ansatz (TBA) equations \cite{YY69} 
 related to $sl(r|s)$. 
 Namely, the TBA equation for the supersymmetric $t-J$ model, 
 which is related to 
 the vector representation of $sl(2|1)$, is proposed in \cite{Sch87}.
  This is also generalized \cite{Sch92}
 to the case of the fundamental representation of $sl(s|1)$. 
 The TBA equation for a supersymmetric extended Hubbard model, 
 which is related to the fundamental representation of $sl(2|2)$, 
 is proposed in \cite{EK94}. 
 One can also find TBA equations beyond the fundamental representations. 
 The TBA equation of a generalized $t-J$ model, which is 
 based on a tensor representation of $sl(2|1)$, 
 is proposed in \cite{Fr99}. 
 As is well known, 
 $sl(r|s)$ admits finite dimensional representations with 
 continuous parameters. 
 In fact the TBA equation of the model 
 for electrons with generalized hopping terms and 
 Hubbard interaction, which is related to one parameter family of 
 four dimensional representations of $sl(2|1)$, 
 is proposed in \cite{BF95}.  
 In the case of $sl(r|s)$, 
 the TBA equation with a continuous parameter is studied \cite{Sa99} 
 in relation with the continuum limit of integrable spin chains. 
 These TBA equations are derived by the traditional 
 string hypothesis \cite{T71,G71}. One can also derive the same 
 TBA equations by the quantum transfer matrix (QTM) method 
\cite{S85,SI87,K87,SAW90,Kl92,JKS98}. 
In fact the TBA equations for the supersymmetric 
$t-J$ model \cite{Sch87} 
and the supersymmetric extended Hubbard model \cite{EK94} are also 
derived \cite{JKS98} by the QTM method based on the 
$sl(r|s)$ version of the $T$-system \cite{T97} 
(a system of functional relations 
 among transfer matrices of fusion models). 

On the other hand, study on TBA equations for $osp(r|2s)$ case 
 has begun quite recently in Refs. \cite{ST99,ST00,ST01}, 
in which we deal with the simplest $osp(1|2)$ model 
 from the point of view of  the string hypothesis 
and the QTM method. 
 The $R$-matrix of the $osp(r|2s)$ models are given as 
 solutions of the graded Yang-Baxter equations \cite{KS82}, which have 
 bosonic and fermionic degree of freedom. 
 The $R$-matrix of the $osp(r|2s)$ models are proposed
  in \cite{Kul86}. 
 Moreover they are constructed 
  from the point of view of automorphisms of Lie superalgebras and 
  classical graded Yang-Baxter equations \cite{BS88}; 
   quantum supergroups \cite{ZBG91}; 
 braid-monoid algebras \cite{MR94}. 
 Some special examples are also examined 
 from various context. See, for example, Refs. \cite{KR89,DFI90,Sa90}. 
 As for eigenvalue formulae of transfer matrices, 
 $osp(1|2)$ model \cite{Kul86} 
 and $osp(1|2s)$ model \cite{Mar95-2} 
 are solved by the analytic Bethe ansatz; 
 $osp(2r-1|2),osp(2|2s-2),osp(2r-2|2),osp(1|2s)$ 
 models \cite{MR97} are solved by the algebraic Bethe ansatz. 
 These works deal with models related to fundamental representations. 
 As for more complicated representations, 
 we have executed an analytic Bethe ansatz \cite{R83,KS95}
  related to  
 $C(s)=osp(2|2s-2)$ \cite{T99-1},
  $B(r|s)=osp(2r+1|2s)$ and $D(r|s)=osp(2r|2s)$ \cite{T99}, 
 in which \symbol{"60}super Yangian analogue of 
 Young supertableaux\symbol{"27} are proposed and 
 $T$-systems among them are also found. 
The $osp(r|2s)$ integrable spin chain is related to interesting physical 
problems, such as the loop model which is 
 related to statistical properties of 
polymers\cite{MNR98}, 
and the fractional quantum Hall effect \cite{HR88}, etc. 
So it is desirable to study the $osp(r|2s)$ integrable spin chain
 beyond the $osp(1|2)$ case. 

The purpose of this paper is to study 
 TBA equations  related to 
 $osp(1|2s)=B(0|s)$ by 
 the string hypotheses \cite{T71,G71} and the 
 QTM method \cite{S85,SI87,K87,SAW90,Kl92,JKS98}. 
 In section 2, 
 we assume a rather general Bethe ansatz equation (BAE), 
 and derive the TBA equation by the string hypothesis. 
 We expect this TBA equation is the one for 
 a higher spin analogue of the $osp(1|2s)$ 
 integrable spin chain in Ref. \cite{Mar95-2}. 
 In fact, the high temperature limit of the free energy 
 is expressed by an appropriate  solution of 
 a difference equation called the $Q$-system. 
 In particular for the 
 fundamental representation case, 
 we also derive the TBA equation from the QTM method in section 4. 
 Namely, we transform the 1D quantum system to the 2D classical system 
 by the Suzuki-Trotter mapping \cite{S85}, and define a QTM, which is 
  a transfer matrix of an inhomogeneous vertex model. 
  We consider the QTM in the context of the fusion hierarchy of the model,
   and derive the functional relations among 
  \symbol{"60}fusion QTMs\symbol{"27} from  
 the $osp(1|2s)$ version of the $T$-system\cite{T99}. 
 After a dependant variable transformation, 
 we obtain the $Y$-system (which is expected to be 
 a system of functional relations 
 among a solution of the TBA equation) 
 from the $T$-system.
  Finally we transform the $Y$-system with certain analytical 
  conditions (ANZC conditions: 
  Analytic NonZero and Constant asymptotics 
  for the large spectral parameter) into the TBA equation.  
 Moreover we find that 
 this TBA equation coincides with the one in section 2. 
 This indicates the validity of the string hypothesis for 
 the $osp(1|2s)$ model. 
 In section 5, we evaluate the 
 low temperature asymptotics of the specific heat. 
 We express the central charge by the Rogers dilogarithmic function 
 and identify it as $c=s$. 
 This coincides with the conjecture \cite{Mar95-2} 
 by the root density method. 
\section{String hypothesis}
For the last several decades, 
many people recognized that 
Bethe ansatz equations (BAE) 
can be written in terms of the representation theoretical 
data of Lie algebras \cite{RW87} or Lie superalgebras \cite{Kul86,MR97}. 
So we assume, as our starting point, 
the following type of the $osp(1|2s)$ Bethe 
ansatz equation on complex variables $\{v_{k}^{(a)}\}$:
\begin{eqnarray}
\prod_{j=1}^{N}\left(
\frac{v^{(a)}_{k}-w^{(a)}_{j}+\frac{i}{2t_{a}}b^{(a)}_{j}}
     {v^{(a)}_{k}-w^{(a)}_{j}-\frac{i}{2t_{a}}b^{(a)}_{j}}
\right)=
-\varepsilon_{a}
\prod_{d=1}^{s+1}
\frac{Q_{\sigma(d)}(v^{(a)}_{k}+\frac{i}{2}B_{ad})}
     {Q_{\sigma(d)}(v^{(a)}_{k}-\frac{i}{2}B_{ad})},
     \label{BAE}
\end{eqnarray}
where $k\in \{1,2, \dots, M_{a}\}$; $a\in \{1,2, \dots, s\}$; 
$\sigma(d)=d$ for $1\le d \le s$; $\sigma(s+1)=s$; 
$t_{a}=1$ for $1\le a \le s-1$; $t_{s}=2$; 
$B_{ad}=2\delta_{ad}-\delta_{a,d+1}-\delta_{a,d-1}$;
$Q_{a}(v)=\prod_{k=1}^{M_{a}}(v-v_{k}^{(a)})$; 
$M_{a}\in {\mathbb Z}_{\ge 0}$; $M_{0}=N$; 
$N$ is the number of the lattice site. 
$w^{(a)}_{j}\in {\mathbb C}$ is an inhomogeneity parameter. 
In this section, we consider the case $w^{(a)}_{j}=0$. 
The parameter $\sigma$ expresses an effect of 
\symbol{"60}a peculiar two-body self-interaction 
 for the root $\{v_{k}^{(s)}\}$\symbol{"27} \cite{MR97}, 
 which originates from the odd simple root $\alpha_{s}$ with 
 $(\alpha_{s}|\alpha_{s})\ne 0$. 
%
\begin{figure}
    \setlength{\unitlength}{0.8pt}
    \begin{center}
    \begin{picture}(250,50) 
      \put(10,20){\circle{20}}
      \put(20,20){\line(1,0){20}}
      \put(50,20){\circle{20}}
      \put(60,20){\line(1,0){20}}
      \put(90,20){\line(1,0){10}}
      \put(110,20){\line(1,0){10}}
      \put(130,20){\line(1,0){20}}
      \put(160,20){\circle{20}}
      \put(170,20){\line(1,0){20}}
      \put(7,-2){$b_{1}$}
      \put(46,-2){$b_{2}$}
      \put(156,-2){$b_{s-2}$}
      \put(200,20){\circle{20}}
      \put(240,20){\circle*{20}}
      \put(208.8,25){\line(1,0){32.4}}
      \put(208.8,15){\line(1,0){32.4}}
      \put(197,-2){$b_{s-1}$}
      \put(236,-2){$b_{s}$}
      \put(230,20){\line(-1,1){14.14214}}
      \put(230,20){\line(-1,-1){14.14214}}
  \end{picture}
  \end{center}
  \caption{The Dynkin diagram for the Lie superalgebra 
  $B(0|s)=osp(1|2s)$ ($s \in {\mathbb Z}_{\ge 1}$):
   white circles denote even roots; 
   a black circle denotes an odd root $\alpha_{s}$ 
   with $(\alpha_{s}|\alpha_{s})\ne 0$. 
   The Kac-Dynkin label $b_{a}$ 
   of an irreducible representation of 
   $osp(1|2s)$ is depicted under each vertex. 
   This irreducible representation is finite dimensional 
   if and only if 
   $b_{a}\in {\mathbb Z}_{\ge 0}$ for $a \in \{1,2,\dots,s-1\}$ 
    and $b_{s}\in 2{\mathbb Z}_{\ge 0}$.}
  \label{dynkin-b0s}
\end{figure}
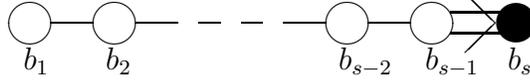
%
The parameters $(b^{(1)}_{j},b^{(2)}_{j},\dots,b^{(s)}_{j})$ 
 in the BAE (\ref{BAE}) represent the 
Kac-Dynkin label \cite{Ka78} 
(see, Figure \ref{dynkin-b0s}). In this paper, we consider the 
case
\begin{eqnarray}
b^{(a)}_{j}=b\delta_{ap}(1-\delta_{ps})+2b\delta_{ap}\delta_{ps} 
\label{Kac-Dynkin},
\end{eqnarray} 
where $b\in {\mathbb Z}_{\ge 1}, p \in \{1,2,\dots,s \}$.
Note that $b^{(s)}_{j}$ is always an even number. 
 We write the irreducible representation of $osp(1|2s)$ 
 labeled by (\ref{Kac-Dynkin}) as $V_{b}^{(p)}$. 
$\varepsilon_{a}$ is a phase factor ($|\varepsilon_{a}|=1$). 
In particular, for $p=b=1$ case, we have \cite{Mar95-2}
\begin{eqnarray}
\varepsilon_{a}=
           \left\{
            \begin{array}{lll}
            (-1)^{N-M_{2}} & {\rm if } & a=1 \\ 
            (-1)^{M_{a-1}-M_{a+1}} 
            & {\rm if } & a \in \{2,3,\dots,s-1\} \\
            (-1)^{M_{s-1}-M_{s}}& {\rm if } & a=s.
            \end{array}
           \right.
\end{eqnarray}
Note that BAE (\ref{BAE}) for $p=b=1$ 
 corresponds to the BAE \cite{Mar95-2,MR97} of 
  the $2s+1$ state $osp(1|2s)$ model \cite{BS88,ZBG91,MR94}
 (the Hamiltonian of this model 
 is given in (\ref{Hamiltonian})).

Next we will present a function $T_{m}^{(a)}(v)$ with a 
spectral parameter $v \in {\mathbb C}$, which 
 is a candidate of an eigenvalue formula (DVF) 
 of a transfer matrix for an $osp(1|2s)$ vertex model of a fusion type 
 for auxiliary space labeled by $(a,m)$. 
 We consider the case where the transfer matrix is defined as the 
 ordinary trace of a monodromy matrix. 
 The auxiliary space $W_{m}^{(a)}$  
 of the corresponding transfer matrix should be   
an irreducible finite dimensional module of the 
\symbol{"60}super Yangian $Y(osp(1|2s))$ \symbol{"27}. 
 As an $osp(1|2s)$-module, 
 $W_{m}^{(a)}$ is expected to contain $V_{m}^{(a)}$ 
 as \symbol{"60}top term\symbol{"27}: any other irreducible component 
 in $W_{m}^{(a)}$ has a highest weight lower than that of $V_{m}^{(a)}$. 
As for $W_{m}^{(a)}$, the multiplicity of the irreducible 
components may be calculated by 
 a Kirillov-Reshetikhin\cite{KR90}-like formula in Ref. \cite{T99}. 
 The quantum space that the transfer matrix acts on 
 is the tensor product module $(W_{b}^{(p)})^{\otimes N}$. 
We can derive an explicit expression of $T_{m}^{(a)}(v)$ 
by modifying the vacuum part of the DVF in Ref. \cite{T99} 
 so that the vacuum part is compatible with the left 
 hand side of the BAE (\ref{BAE}): 
\begin{eqnarray}
T_{m}^{(a)}(v)=\sum_{\{d_{jk}\}} \prod_{j=1}^{m}\prod_{k=1}^{a}
z(d_{jk};v-\frac{i}{2}(m-a-2j+2k)),
\label{DVF}
\end{eqnarray}
where the summation is taken over 
$d_{jk}\in 
\{1,2,\dots,s,0,\overline{s}, \dots, \overline{2},\overline{1}\}$ 
($ 1 \prec 2 \prec \cdots \prec s \prec 0 \prec 
\overline{s} \prec \cdots \prec \overline{2} \prec \overline{1}$)
such that $d_{jk} \preceq d_{j+1,k}$ and $d_{jk} \prec d_{j,k+1}$. 
The functions $\{z(a;v)\}$ are defined as
\begin{eqnarray}
&& z(a;v)=\psi_{a}(v)
\frac{Q_{a-1}(v+\frac{i}{2}(a+1))Q_{a}(v+\frac{i}{2}(a-2))}
{Q_{a-1}(v+\frac{i}{2}(a-1))Q_{a}(v+\frac{i}{2}a)} 
\nonumber \\
&& \hspace{150pt} \mbox{for} \quad a \in \{1,2,\dots,s\}, 
\nonumber \\
&& z(0;v)=\psi_{0}(v)
\frac{Q_{s}(v+\frac{i}{2}(s-1))Q_{s}(v+\frac{i}{2}(s+2))}
     {Q_{s}(v+\frac{i}{2}(s+1))Q_{s}(v+\frac{i}{2}s)}, \\  
&& z(\overline{a};v)=\psi_{\overline{a}}(v)
\frac{Q_{a-1}(v-\frac{i}{2}(a-2s))Q_{a}(v-\frac{i}{2}(a-2s-3))}
{Q_{a-1}(v-\frac{i}{2}(a-2s-2))Q_{a}(v-\frac{i}{2}(a-2s-1))} 
\nonumber \\
&& \hspace{150pt} \mbox{for} \quad a \in \{1,2,\dots,s\},
\nonumber 
\end{eqnarray}
where $Q_{0}(v):=1$.  
The vacuum parts of $\{z(a;v)\}$ are given as follows
\begin{eqnarray}
\psi_{a}(v)=
\left\{
 \begin{array}{l}
 \zeta_{a}  \hspace{97pt} 
 \mbox{for} \quad 1 \preceq a \preceq p ,\\ 
 \zeta_{a}\frac{\phi_{b}(v+\frac{i}{2}(p-1))}{\phi_{b}(v+\frac{i}{2}(p+1))} 
 \hspace{40pt}
 \mbox{for} \quad p+1 \preceq a \preceq \overline{p+1}, \\ 
 \zeta_{a}\frac{\phi_{b}(v+\frac{i}{2}(p-1))\phi_{b}(v-\frac{i}{2}(p-2s))}
      {\phi_{b}(v+\frac{i}{2}(p+1))\phi_{b}(v-\frac{i}{2}(p-2s-2))} 
      \quad
 \mbox{for} \quad \overline{p} \preceq a \preceq \overline{1}, 
 \end{array}
\right.
\label{vac}
\end{eqnarray}
where $\phi_{b}(v)=\prod_{k=1}^{b}(v+\frac{i}{2}(b+1-2k))^{N}$; 
$\zeta_{a}$ is a phase factor ($|\zeta_{a}|=1$): 
$\zeta_{a}=\zeta_{\overline{a}}=
\zeta_{1}\varepsilon_{1}\varepsilon_{2} \cdots \varepsilon_{a-1}$
 for $a \in \{1,2,\dots,s \}$, 
$\zeta_{0}=\zeta_{1}\varepsilon_{1}\varepsilon_{2} \cdots \varepsilon_{s}$. 
In particular for $p=b=1$ case, 
 the phase factor $\zeta_{a}$ is given \cite{Mar95-2} as follows
\begin{eqnarray}
\zeta_{a}=
           \left\{
            \begin{array}{lll}
            (-1)^{N-M_{1}} & {\rm if } & a=1 \\ 
            (-1)^{M_{a-1}-M_{a}} 
            & {\rm if } & a \in \{2,3,\dots,s\} \\
            1& {\rm if } & a=0 \\
            (-1)^{M_{\overline{a}-1}-M_{\overline{a}}} 
            & {\rm if } & 
            a \in \{\overline{s},\dots,\overline{3},\overline{2}\}\\
            (-1)^{N-M_{1}} & {\rm if } & a=\overline{1},
            \end{array}
           \right.
           \label{pf}
\end{eqnarray}
where $\overline{\overline{a}}=a$.
The dress part of (\ref{DVF}) is free of poles under the BAE (\ref{BAE}). 
This is a requirement from the analytic Bethe ansatz \cite{R83,KS95}. 
Note that $T^{(1)}_{1}(v)$ for $p=b=1$ 
 corresponds to the DVF \cite{Mar95-2,MR97} of 
  the $2s+1$ state $osp(1|2s)$ model. 
  We may think of (\ref{DVF}) as $osp(1|2s)$ version of the 
 Bazhanov and Reshetikhin's eigenvalue formula \cite{BR90}. 

We shall consider the one dimensional counterpart of the 
above-mentioned fusion vertex model for the case $(a,m)=(p,b)$,
 which is a higher spin analogue of the $2s+1$ state 
$osp(1|2s)$ spin chain\cite{Mar95-2}. 
The energy density of the corresponding system 
is defined as follows: 
\begin{eqnarray}
\mathcal{E}=\frac{J}{Ni}\frac{d}{dv}
\log T_{b}^{(p)}(v)\biggl|_{v=0}
=\frac{J}{N} \sum_{k=1}^{N_{p}}
\frac{b}{(v_{k}^{(p)})^{2}+(\frac{b}{2})^{2}},
\label{energy}
\end{eqnarray}
where $J$ is a real coupling constant. 
For simplicity, 
the function (\ref{vac}) is normalized so that 
the constant term of the energy 
density (\ref{energy}) vanishes. 
The extra signs (\ref{pf}) do not appear 
in (\ref{energy}) explicitly. 
We adopt the following ordinary string solution for (\ref{BAE}) 
in the thermodynamic limit (cf. Ref. \cite{Mar95}):
\begin{eqnarray}
v_{m,k}^{(a)}+\frac{i}{2}(m+1-2j)
\label{string},
\end{eqnarray}
where $m \in {\mathbb Z}_{\ge 1}$; 
$a \in \{1,2,\dots, s \}$; 
$j \in \{1,2,\dots, m \}$; 
$k \in \{1,2,\dots, n_{m}^{(a)}\}$; 
$v_{m,k}^{(a)}\in \mathbb{R}$; 
$n_{m}^{(a)}$ is the number of color-$a$ $m$-strings. 
Let $\rho_{m}^{(a)}(v)$ and $\sigma_{m}^{(a)}(v)$ 
be string and hole densities for each color $a$ and length $m$. 
Substituting (\ref{string}) into (\ref{BAE}), 
 one can derive a relation between 
 $\rho_{m}^{(a)}(v)$ and $\sigma_{m}^{(a)}(v)$ 
 after some manipulation: 
\begin{eqnarray}
\delta_{ap}\Psi_{m,b}(v)=\sigma_{m}^{(a)}(v)
+\sum_{l=1}^{\infty}\sum_{d=1}^{s} 
\mathbf{A}_{ml}^{ad}\rho_{l}^{(d)}(v), 
\label{constraint}
\end{eqnarray}
where the function $\Psi_{m,b}(v)$ is given by 
\begin{eqnarray}
\Psi_{m,b}(v)=\frac{i}{2\pi}\frac{\partial}{\partial v}
\sum_{j=1}^{m} \log \left\{
\frac{v+\frac{i}{2}(m+1-2j+b)}{v+\frac{i}{2}(m+1-2j-b)}
\right\}
\end{eqnarray}
and the operator $\mathbf{A}_{ml}^{ad}$ is defined by 
\begin{eqnarray}
&& \mathbf{A}_{ml}^{ad}=\delta_{ad}
([|l-m|]+2[|l-m|+2]+\cdots+2[l+m-2]+[l+m]) \nonumber \\
&& \hspace{50pt} -I_{ad}([|l-m|+1]+[|l-m|+3]+\cdots+[l+m-1]),
\nonumber \\
&& \hspace{100pt} 
I_{ad}=\delta_{a,d-1}+\delta_{a,d+1}+\delta_{a d}\delta_{a s}. 
\end{eqnarray}
Here the action of $[m]$ on any function $h(v)$ is 
defined as follows
\begin{eqnarray}
[m]h(v)=\left\{\begin{array}{lll}
         (\Psi_{m,1}*h)(v) & \mbox{if} & m \ne 0 \\ 
         h(v) & \mbox{if} & m = 0,
        \end{array}
        \right.
\end{eqnarray}
where $*$ is a convolution
\begin{eqnarray}
(\Psi_{m,1}*h)(v)=\int_{-\infty}^{\infty}dw \Psi_{m,1}(v-w)h(w).
\end{eqnarray}
In the thermodynamic limit,
 the energy density (\ref{energy}) reduces to 
\begin{eqnarray}
\mathcal{E}=2\pi J \sum_{m=1}^{\infty}
 \int_{-\infty}^{\infty}dv \Psi_{m,b}(v)\rho_{m}^{(p)}(v). 
\end{eqnarray}
The entropy density leads as follows
\begin{eqnarray}
&& \mathcal{S}=k_{B}\sum_{a=1}^{s}\sum_{m=1}^{\infty}
\int_{-\infty}^{\infty}dv 
\left\{(\rho_{m}^{(a)}(v)+\sigma_{m}^{(a)}(v))
\log(\rho_{m}^{(a)}(v)+\sigma_{m}^{(a)}(v)) \right. \nonumber \\
&& \hspace{50pt} \left. -\rho_{m}^{(a)}(v)\log \rho_{m}^{(a)}(v)
-\sigma_{m}^{(a)}(v)\log \sigma_{m}^{(a)}(v)
\right\},
\end{eqnarray}
where $k_{B}$ denotes the Boltzmann constant. 
Now we shall investigate the equilibrium state. 
From the condition 
$\frac{\delta \mathcal{F}}{\delta \rho_{m}^{(a)}(v)}=0$ for 
the free energy density 
$\mathcal{F}=\mathcal{E}-T\mathcal{S}$ ($T$: temperature) 
and the relation (\ref{constraint}), 
we obtain a TBA equation 
\begin{eqnarray}
\log(1+Y_{m}^{(a)}(v))=
2 \pi J \beta \Psi_{m,b}(v)\delta_{ap}
+\sum_{l=1}^{\infty}\sum_{d=1}^{s} 
\mathbf{A}_{ml}^{ad}\log(1+Y_{l}^{(d)}(v)^{-1})
 \label{TBA-1}
\end{eqnarray}
and the free energy density
\begin{eqnarray}
\mathcal{F}=-\frac{1}{\beta}\sum_{m=1}^{\infty}\int_{-\infty}^{\infty}dv 
\Psi_{m,b}(v)
\log(1+(Y_{m}^{(p)}(v))^{-1}), \label{free-en}
\end{eqnarray}
where $Y_{m}^{(a)}(v):=\sigma_{m}^{(a)}(v)/\rho_{m}^{(a)}(v)$
 and $\beta :=1/(k_{B}T)$. 
We shall derive an alternative form of the TBA equation (\ref{TBA-1}). 
Taking note on the relations
\begin{eqnarray}
\hat{\Psi}_{m,b}(k)&=&
\int_{-\infty}^{\infty} dv \Psi_{m,b}(v)e^{-ikv}  \nonumber \\ 
&=& \frac{\sinh (\frac{\min(b,m)k}{2})}{\sinh \frac{k}{2}}
 \exp\left(-\frac{\max(b,m)|k|}{2}\right), \nonumber \\ 
&&\sum_{m=1}^{\infty}\hat{X}_{nm}\hat{\Psi}_{m,b}(k)
=\frac{\delta_{nb}}{2\cosh\frac{k}{2}},\nonumber \\ 
&& \hat{X}_{nm}=\delta_{nm}-
\frac{\delta_{n,m-1}+\delta_{n,m+1}}{2\cosh\frac{k}{2}},
\label{psi}
\end{eqnarray}
we can rewrite (\ref{TBA-1}) as follows: 
\begin{eqnarray}
\log Y_{m}^{(a)}(v)=
\frac{\pi J \beta \delta_{ap}\delta_{mb}}{\cosh\pi v}
+K*\log\left\{
\frac{(1+Y_{m-1}^{(a)})(1+Y_{m+1}^{(a)})}
{\prod_{d=1}^{s}(1+(Y_{m}^{(d)})^{-1})^{I_{ad}}}\right\}(v), 
\label{TBA-2}
\end{eqnarray}
where $a\in \{1,2,\dots,s\}$, $m\in {\mathbb Z}_{\ge 1}$,
 $Y_{0}^{(a)}(v):=0$ and the kernel is 
\begin{eqnarray}
K(v)=\frac{1}{2 \cosh \pi v}. 
\end{eqnarray}
Note that (\ref{TBA-2}) reduces to the TBA equation in 
Refs. \cite{ST99,ST00}, if we set $p=b=s=1$. 
\section{High temperature limit}
We shall consider the high temperature limit of the 
free energy density (\ref{free-en}). 
We assume that the function $Y_{m}^{(a)}(v)$ is  
independent of $v$ in the limit $T \to \infty$. 
In this case, the constant
 solution of (\ref{TBA-2}) obeys the constant $Y$-system
\begin{eqnarray}
(Y_{m}^{(a)})^{2}=
\frac{(1+Y_{m-1}^{(a)})(1+Y_{m+1}^{(a)})}
 {\prod_{d=1}^{s}(1+(Y_{m}^{(d)})^{-1})^{I_{ad}}}, 
 \label{const-Y}
\end{eqnarray}
where $Y_{0}^{(a)}:=0$,
 $a \in \{1,2,\dots, s\}$ and $m \in {\mathbb Z}_{\ge 1}$. 
The solution of this constant $Y$-system is 
\begin{eqnarray}
Y_{m}^{(a)}=
 \frac{Q_{m-1}^{(a)}Q_{m+1}^{(a)}}
      {\prod_{d=1}^{s}(Q_{m}^{(d)})^{I_{ad}}},
      \label{y-Q}
\end{eqnarray}
where $\{ Q_{m}^{(a)} \}$ are the dependant variables of the 
$Q$-system: 
\begin{eqnarray}
(Q_{m}^{(a)})^{2}=Q_{m-1}^{(a)}Q_{m+1}^{(a)}
 +\prod_{d=1}^{s}(Q_{m}^{(d)})^{I_{ad}}
 \label{Q-sys},
\end{eqnarray}
where $Q_{0}^{(a)}:=1$,
 $a \in \{1,2,\dots, s\}$ and $m \in {\mathbb Z}_{\ge 1}$. 
We can derive this $Q$-system by 
neglecting the vacuum part and the spectral parameter dependence 
of the $T$-system in Ref. \cite{T99} (see, (\ref{T-system})): 
a system of fusion relations among commuting transfer matrices 
of the $osp(1|2s)$ vertex model.  
 The $Q$-system was proposed in Ref. \cite{KR90} 
as a system of difference equations among characters of 
  modules of the Yangian $Y(\mathfrak{g})$ 
 as $\mathfrak{g}$ modules 
 ($\mathfrak{g}$: simple Lie algebras).  
Our $Q$-system (\ref{Q-sys}) 
is equivalent to the $A_{2s}^{(2)}$ version of the 
$Q$-system \cite{KS95-2,HKOTT}. 
We think that this coincidence 
originates from the correspondence \cite{Z97} between 
$B^{(1)}(0|s)$ and $A_{2s}^{(2)}$. 
There are several similar correspondences \cite{Z97} 
between superalgebras and ordinary algebras. 
Thus, we expect that 
 \symbol{"60}\symbol{"60}different algebras with the same $Q$-system"
 also occurs to several other algebras. 

We can express the solution of the $Q$-system (\ref{Q-sys}) 
 as a polynomial of the fundamental variables
  $Q_{1}^{(1)},Q_{1}^{(2)},\dots,Q_{1}^{(s)}$. 
  Moreover, this polynomial has a determinant expression: 
\begin{eqnarray}
Q_{m}^{(a)}=\det_{1\leq i,j \leq m}({\cal Q}_{a+i-j}),\label{det}
\end{eqnarray}
where ${\cal Q}_{j}$ satisfies 
\begin{eqnarray}
{\cal Q}_{a}={\cal Q}_{2s-a+1}
\end{eqnarray}
and a condition 
\begin{eqnarray}
{\cal Q}_{a}=\left\{
      \begin{array}{lll}
                  0 & \mbox{for} & a \in {\mathbb Z}_{<0} \\
                  1 & \mbox{for} & a=0 \\
        Q_{1}^{(a)} & \mbox{for} & a \in \{1,2,\dots,s\}. \\ 
      \end{array}
             \right.
\end{eqnarray}
In our case, ${\cal Q}_{a}$ is given as follows 
\begin{eqnarray}
{\cal Q}_{a}=
   \left(
    \begin{array}{c}
    2s+1 \\ 
    a
    \end{array}
   \right). \label{bino}
\end{eqnarray}
In this case, $Q_{m}^{(a)}$ is the number of terms in 
DVF (\ref{DVF}), which is expected \cite{T99} to be the dimension 
 of  $W_{m}^{(a)}$. 
 In particular,
  $Q_{1}^{(1)}$ coincides with the dimension of $V_{1}^{(1)}$. 
  For general case, $Q_{m}^{(a)}$ is 
  greater than or equal to the dimension of $V_{m}^{(a)}$.  
When solving (\ref{TBA-2}), we assume $Y_{m}^{(a)}(\pm \infty)$ 
coincide with the solution (\ref{y-Q}) of the constant Y-system  
with (\ref{det})-(\ref{bino}). 

Substituting (\ref{y-Q}) into (\ref{free-en}), 
we find that the high temperature limit of the entropy density 
is expressed in terms of a constant solution of the $Q$-system (\ref{Q-sys}): 
\begin{eqnarray}
\lim_{T \to \infty} 
\mathcal{S}=
-\lim_{T \to \infty} 
\frac{\mathcal{F}}{T}=
k_{B}\sum_{m=1}^{\infty}\log(1+(Y_{m}^{(p)})^{-1})^{\min(b,m)}
=k_{B}\log Q_{b}^{(p)} .
\end{eqnarray}
In particular for $p=b=1$ case, we have 
\begin{eqnarray}
\lim_{T \to \infty} 
\mathcal{S}=k_{B}\log Q_{1}^{(1)}=k_{B}\log(2s+1).
\end{eqnarray}
This is the logarithm of the dimension of 
$W_{1}^{(1)}$ or $V_{1}^{(1)}$ \cite{Mar95-2}. 
 This result is compatible with the number 
of the state per site of the $osp(1|2s)$ integrable spin chain. 
In closing this section, we note the fact 
that the solution (\ref{y-Q}) of the constant 
$Y$-system (\ref{const-Y}) can also be written as 
\begin{eqnarray}
Y^{(a)}_{m}=\frac{m(g+m)}{a(g-a)},
\label{solution}
\end{eqnarray}
where $g=2s+1$. 
By using (\ref{solution}), one can derive an explicit expression of 
a constant solution of the $Q$-system (\ref{Q-sys})
\begin{eqnarray}
Q^{(a)}_{m}=\left(\frac{(m+g)!m!}{(m+a)!(m+g-a)!}\right)^m
 \prod_{k=1}^{m}\left\{ \frac{(k+a)(k+g-a)}{k(k+g)} \right\}^{k},
\end{eqnarray}
which will provide a dimension formula of the module $W^{(a)}_{m}$.
\section{$T$-system and QTM method}
In this section, we introduce an integrable 
spin chain\cite{Mar95-2,MR97} associated with the 
fundamental representation $W^{(1)}_{1}$, 
and derive the TBA equation from the point of view 
of the QTM method\cite{S85,SI87,K87,SAW90,Kl92,JKS98}. 
A more detailed explanation of the QTM method in the case of 
$osp(1|2)$ can be found in Ref. \cite{ST00}. 

\begin{figure}
\includegraphics[width=0.95\textwidth]{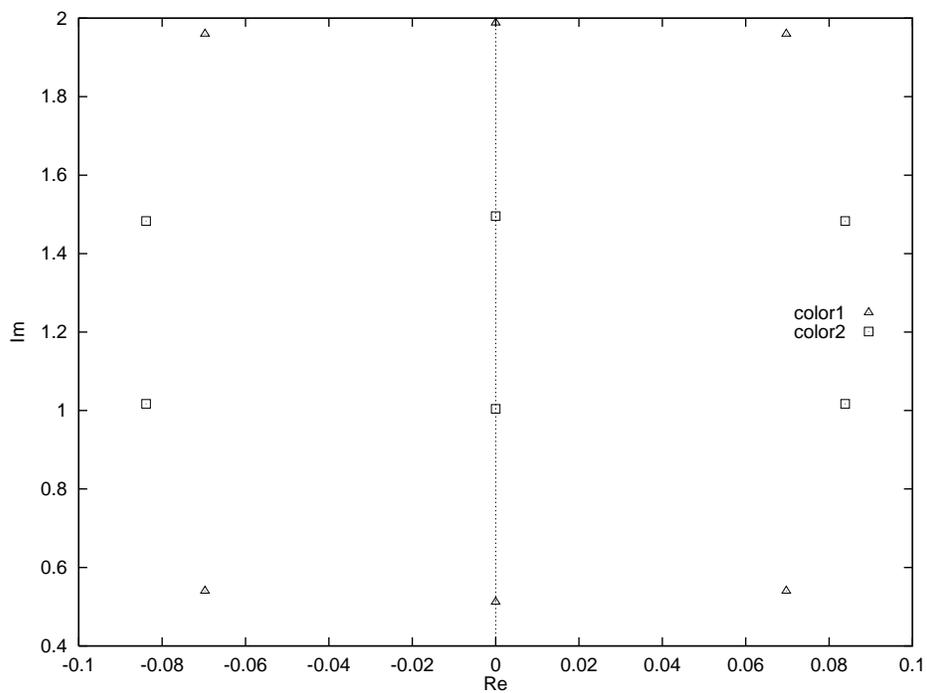}
\caption{Location of the roots of the BAE 
for $osp(1|4)$ case ($N=6$,$u=0.05$), which gives 
the largest eigenvalue of the QTM $T^{(1)}_{1}(v)$. 
Both color 1 roots $\{v^{(1)}_{k}\}$ 
and color 2 roots $\{v^{(2)}_{k}\}$ 
form three two-strings.}
\label{roots}
\end{figure}

\begin{figure}
\includegraphics[width=0.95\textwidth]{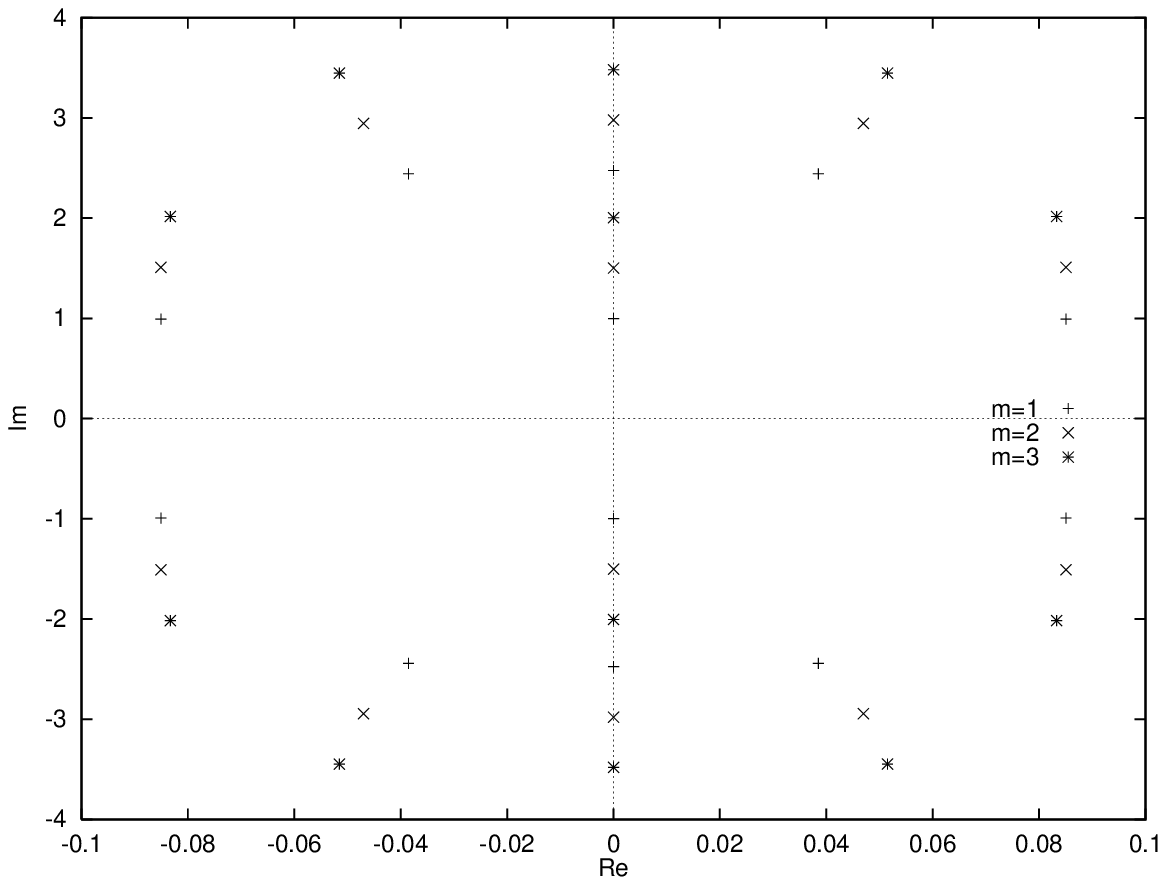}
\caption{Location of zeros of $T^{(1)}_{m}(v)$ 
for $osp(1|4)$ case ($m=1,2,3$, $N=6$,$u=0.05$).
The zeros recede from the physical strip 
${\rm Im}v \in [-\frac{1}{2},\frac{1}{2}]$ as $m$ increases.}
\label{zeros1}
\end{figure}

\begin{figure}
\includegraphics[width=0.95\textwidth]{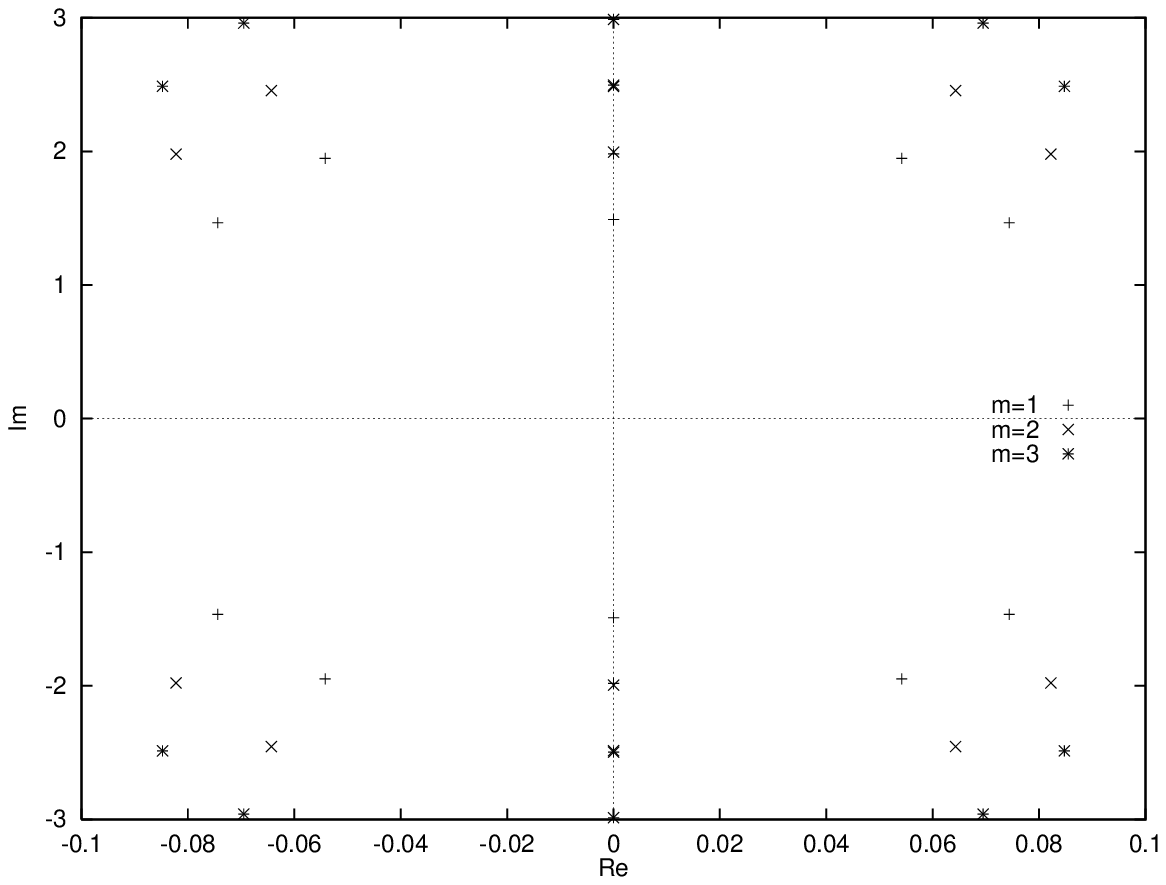}
\caption{Location of zeros of $T^{(2)}_{m}(v)$ 
for $osp(1|4)$ case ($m=1,2,3$, $N=6$,$u=0.05$).
The zeros recede from the physical strip 
${\rm Im}v \in [-\frac{1}{2},\frac{1}{2}]$ as $m$ increases.}
\label{zeros2}
\end{figure}

The $\check{R}$-matrix\cite{BS88,ZBG91,MR94,MR97} of the model is given as 
\begin{eqnarray}
\check{R}(v)=I+vP-\frac{2v}{2v-g}E,
\end{eqnarray}
where $P^{cd}_{ab}=(-1)^{p(a)p(b)}\delta_{ad}\delta_{bc}$; 
$E^{cd}_{ab}=\alpha_{ab}(\alpha^{-1})_{cd}$; 
$a,b,c,d 
\in \{1,2,\dots,s,0,\overline{s},\dots , \overline{1}\} $; 
$\alpha$ is $(2s+1)\times (2s+1)$ anti-diagonal matrix whose 
non-zero elements are  
$\alpha_{a,\overline{a}}=1$ for $a \in \{1,2,\dots,s,0\}$ and  
$\alpha_{a,\overline{a}}=-1$ for 
$a \in \{ \overline{s},\overline{s-1},\dots , \overline{1} \} $; 
$\overline{\overline{a}}=a$; 
$p(a)=0$ for $a=0$; $p(a)=1$ for 
$a \in \{1,2,\dots,s\} \sqcup  
\{\overline{s},\dots,\overline{2},\overline{1}\}$. 
The Hamiltonian of the present 
model for periodic boundary condition is given by 
\begin{eqnarray}
H=J\sum_{k=1}^{L}\left(P_{k,k+1}+\frac{2}{g}E_{k,k+1}\right),
\label{Hamiltonian}
\end{eqnarray}
where $J$ is a coupling constant: $J>0$ and $J<0$ correspond to 
the ferromagnetic and antiferromagnetic regimes respectively; 
$L$ is the number of the lattice sites; $P_{k,k+1}$ and 
$E_{k,k+1}$ act nontrivially 
on the $k$ th site and $k+1$ th site. 
The QTM is defined as 
\begin{eqnarray}
T^{(1)}_{1}(u,v)={\mathrm Tr}_{j}\prod_{k=1}^{\frac{N}{2}}
 R_{a_{2k},j}(u+iv)\widetilde{R}_{a_{2k-1},j}(u-iv),
 \label{QTM}
\end{eqnarray}
where $R^{cd}_{ab}(v)=\check{R}^{cd}_{ba}(v)$; 
 $\widetilde{R}_{jk}(v)=^{t_{k}}\! \! R_{kj}(v)$ 
($t_{k}$ is the transposition in 
the $k$-th space); $N$ is the Trotter number and assumed to even. 
By using the largest eigenvalue $T^{(1)}_{1}(u_{N},0)$ of 
the QTM (\ref{QTM}), 
the free energy density is expressed as 
\begin{eqnarray}
{\mathcal F}=
-\frac{1}{\beta}\lim_{N\to \infty}\log T^{(1)}_{1}(u_{N},0),
\end{eqnarray}
where $u_{N}=-\frac{J\beta}{N}$. 
From now on, we abbreviate the parameter $u$ in 
$T^{(1)}_{1}(u,v)$. 
One can obtain the eigenvalue formulae of the QTM (\ref{QTM}) by 
replacing the vacuum part (\ref{vac}) of the 
DVF (\ref{DVF}) for $p=b=1$ with that of the QTM. 
One may interpret the QTM as a transfer matrix of 
an inhomogeneous vertex model. 
In our case, the inhomogeneity parameters in the BAE (\ref{BAE}) 
for $p=b=1$ 
take the values:
$w^{(a)}_{j}=iu\delta_{a1}$ for $j \in 2{\mathbb Z}_{\ge 1}$; 
$w^{(a)}_{j}=(-iu+\frac{ig}{2})\delta_{a1}$ 
for $j \in 2{\mathbb Z}_{\ge 0}+1$. 
The vacuum parts of (\ref{DVF}) are given as follows
\begin{eqnarray}
\psi_{a}(v)=
\left\{
 \begin{array}{lll}
 \zeta_{1}\frac{\phi_{+}(v)\phi_{-}(v+i)\phi_{+}(v-\frac{2s-1}{2}i)}
   {\phi_{+}(v-\frac{2s+1}{2}i)}  &
 \mbox{for} & a=1,\\ 
 \zeta_{a}\phi_{+}(v)\phi_{-}(v)
 &
 \mbox{for} & 2 \preceq a \preceq \overline{2}, \\ 
 \zeta_{\overline{1}}\frac{\phi_{-}(v)\phi_{+}(v-i)\phi_{-}(v+\frac{2s-1}{2}i)}
   {\phi_{-}(v+\frac{2s+1}{2}i)}
 & \mbox{for} & a=\overline{1}, 
 \end{array}
\right.
\label{vac-QTM}
\end{eqnarray}
where $\phi_{\pm}(v)=(v\pm iu)^{\frac{N}{2}}$; 
 $\zeta_{a}$ is the one in (\ref{pf}). 
For $a \in \{1,2,\dots,s\}$ and $m \in {\mathbb Z}_{\ge 1}$, we 
 define a normalization function
\begin{eqnarray}
{\mathcal N}^{(a)}_{m}(v)=
 \frac{\prod_{j=1}^{m}\prod_{k=1}^{a}
  \phi_{-}(v-\frac{m-a-2j+2k}{2}i)\phi_{+}(v-\frac{m-a-2j+2k}{2}i)}{
  \phi_{-}(v-\frac{m-a}{2}i)\phi_{+}(v+\frac{m-a}{2}i)}.
\end{eqnarray}
 We reset $T^{(a)}_{m}(v)/{\mathcal N}^{(a)}_{m}(v)$ to 
 $T^{(a)}_{m}(v)$, where $T^{(a)}_{m}(v)$ is defined by 
 (\ref{DVF}). 
Since the dress part of the DVF $T^{(a)}_{m}(v)$ is same as  
the row-to-row case, $T^{(a)}_{m}(v)$ satisfies 
the following functional relation, 
which has essentially the same form as the $osp(1|2s)$ 
$T$-system in Ref.\cite{T99}. 
\begin{eqnarray}
&& \hspace{-20pt}
T^{(a)}_{m}(v+\frac{i}{2})T^{(a)}_{m}(v-\frac{i}{2})
=T^{(a)}_{m+1}(v)T^{(a)}_{m-1}(v)+T^{(a-1)}_{m}(v)T^{(a+1)}_{m}(v)
\nonumber \\ 
&& \hspace{130pt} {\rm for} \quad a \in {1,2,\dots,s-1}, 
\label{T-system} \\ 
&& \hspace{-20pt}
T^{(s)}_{m}(v+\frac{i}{2})T^{(s)}_{m}(v-\frac{i}{2})
=T^{(s)}_{m+1}(v)T^{(s)}_{m-1}(v)+
g^{(s)}_{m}(v)T^{(s-1)}_{m}(v)T^{(s)}_{m}(v), \nonumber 
\end{eqnarray}
where 
\begin{eqnarray}
T^{(a)}_{0}(v)&=&\phi_{-}(v+\frac{a}{2}i)\phi_{+}(v-\frac{a}{2}i) 
\quad {\rm for} \quad a \in {\mathbb Z}_{\ge 1},\nonumber \\
T^{(0)}_{m}(v)&=&\phi_{-}(v-\frac{m}{2}i)\phi_{+}(v+\frac{m}{2}i)
\quad {\rm for} \quad m \in {\mathbb Z}_{\ge 1}, \\
g^{(s)}_{m}(v)&=&
 \frac{\phi_{-}(v+\frac{m+s+1}{2}i)\phi_{+}(v-\frac{m+s+1}{2}i)}
 {\phi_{-}(v+\frac{m+s}{2}i)\phi_{+}(v-\frac{m+s}{2}i)}
\quad {\rm for} \quad m \in {\mathbb Z}_{\ge 1}.\nonumber 
\end{eqnarray}
For $s=1$, $g^{(1)}_{m}(v)T^{(0)}_{m}(v)$ coincides with 
the function $T^{(0)}_{m}(v)$ in Ref. \cite{ST00}.
For $m \in {\mathbb Z}_{\ge 1}$, we define the $Y$-functions:
\begin{eqnarray}
 Y^{(a)}_{m}(v)&=&
 \frac{T^{(a)}_{m+1}(v)T^{(a)}_{m-1}(v)}{T^{(a-1)}_{m}(v)T^{(a+1)}_{m}(v)}
 \quad {\rm for} 
 \quad a \in \{1,2,\dots,s-1\}, \nonumber \\
 Y^{(s)}_{m}(v)&=&
 \frac{T^{(s)}_{m+1}(v)T^{(s)}_{m-1}(v)}
 {g^{(s)}_{m}(v)T^{(s-1)}_{m}(v)T^{(s)}_{m}(v)}.
 \label{Y-fun}
\end{eqnarray}
By using the $T$-system (\ref{T-system}), one can show that 
the $Y$-functions satisfy the following $Y$-system:
\begin{eqnarray}
 Y^{(a)}_{m}(v+\frac{i}{2})Y^{(a)}_{m}(v-\frac{i}{2})&=&
 \frac{(1+Y^{(a)}_{m+1}(v))(1+Y^{(a)}_{m-1}(v))}
 {\prod_{d=1}^{s}(1+(Y_{m}^{(d)}(v))^{-1})^{I_{ad}}},
 \label{Y-sys}
 \end{eqnarray}
where $Y^{(a)}_{0}(v)=0$, $a \in \{1,2,\dots,s\}$ and 
$m \in {\mathbb Z}_{\ge 1}$. 
A numerical analysis for finite $N,u,s$ indicates that 
a two-string solution (for every color) in the sector 
$N=M_{1}=M_{2}=\cdots =M_{s}$ of the BAE (\ref{BAE}) provides 
the largest eigenvalue of the QTM (\ref{QTM}) at $v=0$.
Moreover, we expect the following conjecture is valid for 
this two-string solution (see Figure \ref{roots}, 
Figure \ref{zeros1}, Figure \ref{zeros2}).
\begin{conjecture}
For small $u$ ($|u|\ll 1$) and $a \in \{1,2,\dots,s\}$,
every zero of $T^{(a)}_{m}(v)$ is located outside of 
the physical strip 
${\rm Im}v \in [-\frac{1}{2},\frac{1}{2}]$.
\end{conjecture}
Based on this conjecture, 
we shall establish the ANZC property in some domain 
for the $Y$-functions (\ref{Y-fun}) 
to transform the $Y$-system (\ref{Y-sys}) 
 to nonlinear integral equations. 
Here ANZC means Analytic NonZero and Constant asymptotics in the 
limit $|v| \to \infty$. 
One can show that the $Y$-function has the following asymptotic value 
\begin{eqnarray}
\lim_{|v| \to \infty}Y^{(a)}_{m}(v)=\frac{m(g+m)}{a(g-a)},
\label{limit}
\end{eqnarray}
which is identified to the solution (\ref{solution}) 
of the constant $Y$-system (\ref{const-Y}). 
From the conjecture and (\ref{limit}), we find that the 
functions $1+Y_{m}^{(a)}(v)$, $1+(Y_{m}^{(a)}(v))^{-1}$ 
 in the domain ${\rm Im}v \in [-\delta,\delta]$ 
($0<\delta \ll 1$) and  
$Y^{(a)}_{m}(v)$ for $(a,m)\ne (1,1)$ in the domain  
${\rm Im}v \in [-\frac{1}{2},\frac{1}{2}]$ (physical strip) 
have the ANZC property. On the other hand, 
$Y^{(1)}_{1}(v)$ has zeros of order $N/2$ 
at $\pm i(\frac{1}{2}-u)$ if $u>0$ ($J<0$), 
poles of order $N/2$ 
at $\pm i(\frac{1}{2}+u)$ if $u<0$ ($J>0$) in the physical strip. 
Then we must modify $Y^{(1)}_{1}(v)$ as 
\begin{eqnarray}
\hspace{-10pt}
\widetilde{Y}^{(a)}_{m}(v)=Y^{(a)}_{m}(v)\left\{
\tanh\frac{\pi}{2}(v+i(\frac{1}{2} \pm u))\tanh\frac{\pi}{2}
(v-i(\frac{1}{2} \pm u))\right\}^{\pm \frac{N\delta_{a1}\delta_{m1}}{2}}
\hspace{-20pt} ,
\end{eqnarray}
where the sign $\pm $ is identical to that of $-u$. 
Taking note on the relation
\begin{eqnarray}
\tanh\frac{\pi}{4}(v+i)\tanh\frac{\pi}{4}(v-i)=1,
\end{eqnarray}
one can modify the lhs of the $Y$-system (\ref{Y-sys}) as
\begin{eqnarray} 
&& \widetilde{Y}^{(a)}_{m}(v-\frac{i}{2})\widetilde{Y}^{(a)}_{m}(v+\frac{i}{2})
 =\frac{(1+Y^{(a)}_{m+1}(v))(1+Y^{(a)}_{m-1}(v))}
 {\prod_{d=1}^{s}(1+(Y_{m}^{(d)}(v))^{-1})^{I_{ad}}}, 
 \label{modi-Y-sys} \\
&& \hspace{120pt}{\rm for} \quad m \in {\mathbb Z}_{\ge 1} 
\quad {\rm and} \quad a\in \{1,2,\dots,s\}.\nonumber 
\end{eqnarray}
Now that the ANZC property has been established for the $Y$-system, 
we can transform (\ref{modi-Y-sys}) into a system of 
nonlinear integral equations 
by a standard procedure. 
\begin{eqnarray}
\log Y_{m}^{(a)}(v)&=&
\mp \frac{N\delta_{a1}\delta_{m1}}{2}
\log \left\{
\tanh\frac{\pi}{2}(v+i(\frac{1}{2} \pm u))\tanh\frac{\pi}{2}
(v-i(\frac{1}{2} \pm u))
\right\} \nonumber \\
&& +K*\log\left\{
\frac{(1+Y_{m-1}^{(a)})(1+Y_{m+1}^{(a)})}
{\prod_{d=1}^{s}(1+(Y_{m}^{(d)})^{-1})^{I_{ad}}}\right\}(v), 
\label{nonlinear}
\end{eqnarray}
where $Y^{(a)}_{0}(v)=0$, $a\in \{1,2,\dots,s\}$ and
 $m \in {\mathbb Z}_{\ge 1}$.
Substituting $u=-\frac{\beta J}{N}$ and 
taking the Trotter limit $N \to \infty$, 
we obtain the TBA equation 
 (\ref{TBA-2}) for $p=b=1$. 
Taking note on the relations
\begin{eqnarray}
&& C_{ad}(v)=\sum_{l=1}^{\min(a,d)}G_{|a-d|+2l-1}(v) ,\nonumber \\
&& G_{a}(v)=\frac{4}{2s+1}
\frac{\cos\frac{(2s+1-2a)\pi}{4s+2} \cosh\frac{2\pi v}{2s+1}}
{\cos\frac{(2s+1-2a)\pi}{2s+1} + \cosh\frac{4\pi v}{2s+1}},\nonumber \\
&& \widehat{C}_{ad}(k)=\int_{-\infty}^{\infty}{\mathrm d}v
 C_{ad}(v)e^{-ikv}, \nonumber \\
&& \sum_{c=1}^{s}\widehat{C}_{ac}(k)\widehat{D}_{cd}(k)=\delta_{ad},
\label{relations} \\
&& \widehat{D}_{cd}(k)=2\delta_{cd}\cosh\frac{k}{2}-I_{cd},\nonumber
\end{eqnarray}
 one can also rewrite this TBA equation as
\begin{eqnarray}
\log Y_{m}^{(a)}(v)&=& 2\pi \beta J \delta_{m1}G_{a}(v) \nonumber \\ 
&& +\sum_{j=1}^{s}C_{aj}*\log\left\{
\frac{(1+Y_{m-1}^{(j)})(1+Y_{m+1}^{(j)})}
{\prod_{d=1}^{s}(1+Y_{m}^{(d)})^{I_{jd}}}\right\}(v),
\label{TBA3}
\end{eqnarray}
where $Y^{(a)}_{0}(v)=0$, $a\in \{1,2,\dots,s\}$ and
 $m \in {\mathbb Z}_{\ge 1}$.
In contrast to (\ref{TBA-2}), (\ref{TBA3}) does not 
contain $1+(Y^{(a)}_{m}(v))^{-1}$ which is not relevant to 
evaluate the central charge  for the case $J<0$ 
(see, the next section). 
One can derive the following relation 
from (\ref{T-system}) for $a=1$, (\ref{Y-fun}) and (\ref{psi}).
\begin{eqnarray}
\hspace{-23pt} 
\log T^{(1)}_{1}(v)&=&\log \phi_{-}(v+i)\phi_{+}(v-i) \nonumber \\ 
 && +\sum_{m=1}^{\infty}\Psi_{1,m}*\log(1+(Y^{(1)}_{m})^{-1})(v) .
 \label{kitarou}
\end{eqnarray}
Taking the Trotter limit $N \to \infty$ with $u=-\frac{J\beta}{N}$, 
 we find that 
 ${\mathcal F}=-\frac{1}{\beta}\log T^{(1)}_{1}(0)$ 
 with (\ref{kitarou}) coincides with 
the free energy density (\ref{free-en}) for $p=b=1$ 
(if we neglect the first term in the rhs 
of (\ref{kitarou}) coming from a different 
normalization of the DVF). 
One can also derive the following relation 
from (\ref{T-system}) for $m=1$, (\ref{Y-fun}) and (\ref{relations}).
\begin{eqnarray}
\log T^{(1)}_{1}(v)&=&\log \phi_{-}(v+i)\phi_{+}(v-i) 
 +\sum_{a=1}^{s}G_{a}*\log(1+Y^{(a)}_{1}) 
 \nonumber \\
&& +N \int_{0}^{\infty}{\mathrm d}k 
 \frac{2e^{-\frac{k}{2}}\sinh(ku)\cos(kv)\cosh(\frac{2s-1}{4}k)}{
 k \cosh(\frac{2s+1}{4}k)}
\end{eqnarray}
Taking the Trotter limit $N \to \infty$ with $u=-\frac{J\beta}{N}$, 
we obtain 
the free energy density without infinite sum. 
\begin{eqnarray}
{\mathcal F}&=&J\left\{
\frac{2}{2s+1}\left(
2\log 2 -\psi(\frac{1}{2s+1})+\psi(\frac{3+2s}{2+4s})
\right)
-1 
\right\} \nonumber \\
&&-k_BT
\sum_{a=1}^{s}
\int_{-\infty}^
{\infty}{\mathrm d}v G_{a}(v)\log(1+Y^{(a)}_{1}(v)), 
\label{free-finite}
\end{eqnarray}
where $\psi(z)$ is the digamma function
\begin{eqnarray}
\psi(z)&=&\frac{d}{dz}\log \Gamma (z).
\end{eqnarray}
The first term in the rhs of (\ref{free-finite}) for $J=-1$ 
coincides with the grand state energy of
 the $osp(1|2s)$ model in \cite{Mar95-2}.
\section{Central charge}
It was conjectured \cite{Mar95-2} 
that the $osp(1|2s)$ model based on the 
fundamental representation $W^{(1)}_{1}$ is governed 
by the $c=s$ conformal field theory. 
We shall evaluate the central charge for the case 
$p=b=1$ by using the 
result in the previous section and 
 technique in \cite{KP92,KNS94-2}. 
In this section, we assume $J=-1$ and $k_{B}=1$. 
We define a scaling function in the low temperature limit
\begin{eqnarray}
y^{(a)}_{m,\pm}(v)=\lim_{T \to 0}
 Y^{(a)}_{m}(\pm(v +\frac{1}{v_{F}}\log \frac{\gamma}{T})),
 \label{s-fun}
\end{eqnarray}
where $\gamma $ is a constant and $v_{F}$ is the Fermi velocity
\cite{Mar95-2}: $v_{F}=\frac{2\pi}{g}$. 
As $y^{(a)}_{m,+}(v)$ and $y^{(a)}_{m,-}(v)$ 
behave in the same way, we set
 $y^{(a)}_{m}(v)=y^{(a)}_{m,\pm}(v)$. 
From the TBA equation (\ref{TBA3}), 
$y^{(a)}_{m}(v)$ satisfies the following 
equation
\begin{eqnarray}
\log y_{m}^{(a)}(v)&=& -\frac{4v_{F}\delta_{m1}}{\gamma }
{\mathrm e}^{-v_{F}v} \sin (\frac{v_{F}a}{2}) \nonumber \\
&& +\sum_{j=1}^{s}C_{aj}*\log\left\{
\frac{(1+y_{m-1}^{(j)})(1+y_{m+1}^{(j)})}
{\prod_{d=1}^{s}(1+y_{m}^{(d)})^{I_{jd}}}\right\}(v),
\label{TBA4}
\end{eqnarray}
where $y^{(a)}_{0}(v):=0$, $a\in \{1,2,\dots,s \}$ and 
$m\in {\mathbb Z}_{\ge 1}$. 
The first term in rhs of (\ref{TBA4}) is proportional to 
the $a$-th component $\tau_{a}$ of the 
Perron-Frobenius eigenvector of $I_{ac}$: 
\begin{eqnarray}
&& \sum_{c=1}^{s}I_{ac}\tau_{c}=
\left(2\cos \frac{\pi}{g} \right)\tau_{a} 
 \quad {\rm for} \quad a \in \{1,2,\dots,s\},\nonumber\\
&& \tau_{a}=\sqrt{\frac{4}{g}}\sin(\frac{\pi a}{g}), 
\qquad \sum_{a=1}^{s}\tau_{a}^2=1. 
\end{eqnarray}
We assume the DVF $T^{(1)}_{1}(v)$ is expressed as a 
product of the ground state part $T^{(1)gs}_{1}(v)$ 
and the finite temperature correction part $T^{(1)fn}_{1}(v)$. 
In particular we can express the finite size correction part 
 for small $T$ by the function (\ref{s-fun}).
\begin{eqnarray}
&& \log T^{(1)fn}_{1}(v)=\sum_{a=1}^{s}G_{a}*\log (1+Y^{(a)}_{1})(v)
 \\
&& \hspace{10pt} = \frac{4v_{F}T}{\pi \gamma}\cosh (v_{F} v)
 \sum_{a=1}^{s}\sin (\frac{v_{F}a}{2})
 \int_{-\infty}^{\infty}{\mathrm d} w {\mathrm e}^{-v_{F}w}
 \log (1+y^{(a)}_{1}(w)) + o(T) \nonumber .
\end{eqnarray}
Using the relation $G_{a}(-v)=G_{a}(v)$, one can 
derive the following relation from (\ref{TBA4}).
\begin{eqnarray}
&& \frac{4v_{F}^{2}}{\gamma}
\sum_{a=1}^{s}\sin (\frac{v_{F}a}{2})
 \int_{-\infty}^{\infty}{\mathrm d} w {\mathrm e}^{-v_{F}w}
 \log (1+y^{(a)}_{1}(w)) \nonumber \\
&& \hspace{20pt} =-\sum_{a=1}^{s}\sum_{m=1}^{\infty}
 \left\{ L\left( \frac{1}{1+y^{(a)}_{m}(\infty)}\right)
 -L\left( \frac{1}{1+y^{(a)}_{m}(-\infty)}\right)
 \right\},
\end{eqnarray}
where $L(x)$ is the Rogers dilogarithmic function
\begin{eqnarray}
L(x)=-\frac{1}{2}\int_{0}^{x} 
{\mathrm d}y\left\{\frac{\log y}{1-y} +\frac{\log (1-y)}{y} \right\}
\quad {\rm for} \quad 0 \le x \le 1.
\end{eqnarray}
For small $T$, 
the leading term of the specific heat 
\begin{eqnarray}
C=\frac{\partial}{\partial T}
\left( T^{2}\frac{\partial}{\partial T}
 \log T^{(1)fn}_{1}(0) \right)
\end{eqnarray}
is proportional\cite{BCN86,A86} to the central charge c 
\begin{eqnarray}
C=\frac{\pi c T}{3v_{F}} +o(T).
\end{eqnarray}
Then we obtain
\begin{eqnarray}
c=-\frac{6}{\pi^2}\sum_{a=1}^{s}\sum_{m=1}^{\infty}
 \left\{ L\left( \frac{1}{1+y^{(a)}_{m}(\infty)}\right)
 -L\left( \frac{1}{1+y^{(a)}_{m}(-\infty)}\right)
 \right\}. \label{center}
\end{eqnarray}
This expression is independent of the choices of the 
constant $\gamma $ and widely seen in the model related to 
rank $s$ algebras\cite{K93,KNS94-2}. 
To proceed further, 
we have to evaluate the limit $y^{(a)}_{m}(\pm \infty)$.
At first, we expect $\lim_{v \to \infty}y^{(a)}_{m}(v)$ is given 
by (\ref{limit}). 
On the other hand, the divergence from the first term 
in rhs of (\ref{TBA4}) in the limit $v \to -\infty $ 
is expected to be canceled 
by lhs if 
\begin{eqnarray}
 y^{(a)}_{1}(v) \to +0 
 \quad {\rm for } \quad v \to -\infty .
\end{eqnarray}
Then  (\ref{TBA4}) in the limit $v \to -\infty $ 
will be valid if 
$y^{(a)}_{m}(-\infty)$ for $m \in {\mathbb Z}_{\ge 2}$ 
obey the constant $Y$-system (\ref{const-Y}). 
Thus we expect $y^{(a)}_{m}(-\infty)$ is given as the solution 
(\ref{solution}) 
of (\ref{const-Y}) with a shift $m \to m-1$:
\begin{eqnarray}
\lim_{v \to -\infty}y^{(a)}_{m}(v)=\frac{(m-1)(g+m-1)}{a(g-a)}.
\label{limit2}
\end{eqnarray}
Using the relations (\ref{center}), (\ref{limit}), 
(\ref{limit2}) and $L(1)=\frac{\pi^2}{6}$, 
$L(0)=0$, we find $c=s$. 
This result is consistent with the conjecture \cite{Mar95-2} 
by the root density method. 
\section{Discussion}
In this paper, we have proposed the TBA equation for $osp(1|2s)$. 
It will be an interesting future problem to study the TBA equation 
based on a more general orthosymplectic Lie superalgebra $osp(r|2s)$ 
($r>1$). 
As for the string hypothesis, 
we will need to consider  
 more complicated strings than (\ref{string}) 
  as $sl(r|s)$ case\cite{Sch87,Sch92,EK94,Fr99,Sa99}. 
 On the other hand, 
 we have only the $T$-system for tensor-like representations \cite{T99}. 
 To construct complete set of the $T$-system which is relevant for 
 the QTM method, we have to treat spinorial representations.

QTMs  which act on ${\mathbb Z}_{2}$ graded vector space 
 are proposed in \cite{SSSU,LF01}. 
 In particular in \cite{LF01}, two types of QTMs, 
 which give the same partition function in the 
 Trotter limit $N \to \infty $ are proposed. 
 The one is defined as the supertrace of a monodromy matrix 
 of an inhomogeneous graded vertex model. In this case, 
 the sign of the BAE (\ref{BAE}) will have the form 
$\varepsilon_{a}=(-1)^{{\rm deg}(\alpha_{a})}$ ( 
${\rm deg}(\alpha_{a})=0$ for the even roots; 
${\rm deg}(\alpha_{s})=1$ for the odd root). 
The other is defined as the ordinary trace of the monodromy matrix. 
In this case, the sign of the BAE (\ref{BAE}) will disappear 
($\varepsilon_{a}=1$). 
As far as the ground state case is concerned, the difference among 
these formulation of the QTM and the 
ones in this paper may not be important. 

A fermionic formula for $A^{(2)}_{2s}$ coming from a 
different labeling of the Dynkin diagram is proposed in 
\cite{OSS}. Taking account of the correspondence \cite{Z97} between 
$A^{(2)}_{2s}$ and $B^{(1)}(0|s)$, this formula 
may also be relevant to $osp(1|2s)$ case. 
\section*{Acknowledgments}
\noindent
The author would like to thank  
 K. Sakai for collaboration in the previous works 
 \cite{ST99,ST00,ST01}. 
\section*{Note Added}
After the submission of this paper, 
a preprint \cite{SK01} appears, where TBA equations 
related to $osp(1|2s)$ models 
are discussed from the point of view of 
integrable field theories. 
              

\end{document}